# Superconducting-transition-temperature dependence of superfluid density and conductivity in pressurized cuprate superconductors

Jinyu Zhao[1,4]*, Shu Cai [2]*, Yiwen Chen[1,4]*, Genda Gu[3], Hongtao Yan[1], Jing Guo[1], Jinyu Han[1,4], Pengyu Wang[1,4], Yazhou Zhou[1], Yanchun Li[5], Xiaodong Li[5], Zhian Ren[1], Qi Wu[1], Xingjiang Zhou[1,4], Yang Ding[2], Tao Xiang[1,4,6], Ho-kwang Mao[2] and Liling Sun[1,2,4]†

[1]*Institute of Physics, Chinese Academy of Sciences, Beijing 100190, China*
[2]*Center for High Pressure Science & Technology Advanced Research, 100094 Beijing, China*
[3]*Condensed Matter Physics & Materials Science Department, Brookhaven National Laboratory, NY, 11973-5000 USA*
[4]*University of Chinese Academy of Sciences, Beijing 100190, China*
[5]*Institute of High Energy Physics, Chinese Academy of Science, Beijing 100049, China*
[6]*Beijing Academy of Quantum Information Sciences, Beijing 100193, China*

What factors fundamentally determine the value of superconducting transition temperature ($T_c$) in high temperature superconductors has been the subject of intense debate. Following the establishment of an empirical law known as Homes' law, there is a growing consensus in the community that the $T_c$ value of the cuprate superconductors is closely linked to the superfluid density ($\rho_s$) of its ground state and the conductivity ($\sigma$) of its normal state. However, all the data supporting this empirical law ($\rho_s = A\sigma T_c$) have been obtained from the ambient-pressure superconductors. In this study, we present the first high-pressure results about the connection of the quantities of $\rho_s$ and $\sigma$ with $T_c$, through the studies on the $Bi_{1.74}Pb_{0.38}Sr_{1.88}CuO_{6+\delta}$ and $Bi_2Sr_2CaCu_2O_{8+\delta}$, in which the value of their high-pressure resistivity ($\rho=1/\sigma$) is achieved by adopting our newly established method, while the quantity of $\rho_s$ is extracted using the Homes' law. We highlight that the $T_c$ values are strongly linked to the joint response factors of magnetic field and electric field, *i.e.* $\rho_s$ and $\sigma$, respectively, implying that the physics determining $T_c$ is governed by the intrinsic electromagnetic fields of the system.

There are more than two hundred copper-oxide (cuprate) superconductors have been found after the discovery of superconductivity in Ba-doped $La_2CuO_4$[1-3], in which their $T_c$ values vary widely from 35 K in Ba-doped $La_2CuO_4$ to 134 K in $HgBa_2Ca_2Cu_3O_9$ at ambient pressure[4], and to 164 K under the pressure of 30 GPa[5]. Despite intense investigations have been done in the past decades, the issues about why the critical transition temperature ($T_c$) is high for some cuprate superconductors and low for some others, why $T_c$ can be tuned by the parameters such as doping, pressure or magnetic field, and what the physical quantities decisively control the superconducting properties, are still on the list of mysteries of the condensed mater physics[6-25]. Previous studies found that $T_c$ of the superconductors is associated with the superfluid density (or penetration depth) and conductivity (or effective mass)[26-28], however these empirical relations work well only for some of the high-$T_c$ superconductors. Based on the these relations, a more powerful empirical law was proposed by Homes *et al* (known as Homes' law)[11], which states that the superfluid density ($\rho_s$) in the ground state and the normal-state conductivity ($\sigma$) just above $T_c$ are directly connected to the $T_c$ value of both conventional and high-temperature superconductors - $\rho_s = A\sigma \times T_c$ ($A$ is a constant). Currently, Homes' law can provide reasonable description for many ambient-pressure superconducting materials, however, the validity of this law for superconductors under high pressure remains largely unknown. If we assume that Homes' law is applicable to all bulk superconductors, it should hold true for the superconductivity that can be tuned through various methods. To know the connection of the superconductivity in compressed high-$T_c$ superconducting materials with superfluid density and normal-state conductivity, we performed high-pressure resistance measurements on the bulk cuprate superconductors that contain single and double $CuO_2$ planes in a unit cell. Subsequently, we utilized our newly established method to convert the measured resistance under pressure into resistivity[29], enabling us to examine the correlation among the three physical quantities, $T_c$, $\sigma$ and $\rho_s$, with increasing pressure.

Figure 1 illustrates the temperature dependence of *in-plane* resistance for the three over-doped $Bi_{1.74}Pb_{0.38}Sr_{1.88}CuO_{6+\delta}$ (referred to as Bi-2201) single crystals, measured at different pressures. As shown in Fig.1a-1c, these samples display similar superconducting transition temperatures (18 K, 19 K, and 21 K, respectively) at ambient pressure. Furthermore, we observed that the normal-state resistance of these samples initially decreases until it reaches a minimum value. For Sample #1, this minimum resistance occurs at 4.1 GPa, for Sample #2 at 4.0 GPa, and for Sample #3 at 3.8 GPa (Fig. 1a-1c). However, beyond these pressure points, the normal-state resistance increases significantly with further compression. To determine the normal-state conductivity ($\sigma$) for the samples, we employed our newly established method[29] to convert the measured resistance values to resistivity ($\rho$). Subsequently, we obtained $\sigma$ by taking the reciprocal of $\rho$ ($\sigma = 1/\rho$). The obtained resistivity versus temperature were then plotted in Fig. 1d-1f.

It is generally believed that the stability of superconductivity is usually related to the crystal structure. To investigate the structure stability of the compressed sample, we conducted the high-pressure synchrotron X-ray diffraction measurements on the

$Bi_{1.74}Pb_{0.38}Sr_{1.88}CuO_{6+\delta}$ sample at 300 K at 4W2 beamline of the Beijing Synchrotron Radiation Facility. As shown in Fig.2a, all the peaks observed in the X-ray diffraction measurements can be precisely indexed by the orthorhombic structure in space group *Pnan*. Notably, no new peaks were detected within the pressure range of up to 20.5 GPa. Based on the analysis results, we obtained the lattice parameters and volume as a function of pressure (Fig. 2b-2d). It is seen that the lattice parameters and volume consistently decrease as the pressure increases, indicating that no structural phase transition occurs within the pressure range investigated. This lack of structural phase transition is essential for the successful implementation of our method to convert resistance to resistivity[29].

We summarize our high-pressure results of the coevolution of $T_c$, $\sigma$, and the value of $\rho_s$ calculated through Homs' law for the over-doped Bi-2201 in the pressure-$T_c$ phase diagram (Fig. 3a-3c), together with the results achieved from the over-doped Bi-2212 sample (Fig. 3d-3f) for comparison. To assess the validity of the $\rho_s$ value obtained using our method, we performed extrapolation of our high-pressure data ($\rho_s(P)$) to ambient pressure and compared the extrapolated ambient-pressure $\rho_s$ values ($1.5 \times 10^7$ cm$^{-2}$ for Bi-2201 and $9.4 \times 10^7$ cm$^{-2}$ for Bi-2212) with the ambient-pressure $\rho_s$ values ($1.2 \times 10^7$ cm$^{-2}$ for Bi-2201 and $9.3 \times 10^7$ cm$^{-2}$ for Bi-2212) reported for these two types of materials[30,31] (see red diamond and circle in Fig.3c and 3f). Encouragingly, we found that they exhibit good agreement, indicating the reliability of the $\rho_s(P)$ obtained from our high-pressure investigation.

From the results observed at ambient pressure, we note that the average $T_c$ value of the three Bi-2201 samples is about 19.3 K (Fig.3a), which is significantly lower than that of the Bi-2212 sample (Fig.3d). The reason behind the higher $T_c$ value in Bi-2212 compared to Bi-2201 still lacks a comprehensive explanation based on experimental data, although some theoretical investigations proposed that this difference is primarily attributed to the coupling between the layers[32-34]. In order to unravel this puzzle, we conducted estimations of the pressure dependence of conductivity ($\sigma$) and superfluid density ($\rho_s$) for these two different kinds of materials (Fig.3b-3c, Fig.3e-3f and SI). Interestingly, we found that although the ambient-pressure $\sigma$ values at $T_c$ for both Bi-2201 and Bi-2212 are basically at the same level (Fig.3b and 3e), but the $\rho_s$ value of the Bi-2201 samples is nearly 10 times lower than that of Bi-2212 (Fig.3c and 3f), suggesting that the superfluid density ($\rho_s$) plays a crucial role in determining the $T_c$ values of these two high-$T_c$ superconductors.

Furthermore, we also found that the superconducting transition temperatures ($T_c$) of these two kinds of superconductors exhibit low sensitivity to the pressure applied in the experimental range investigated (Fig.3a and 3d). In contrast, both the conductivity ($\sigma$) and superfluid density ($\rho_s$) display significant changes as the pressure increases, exhibiting a dome-like shape (Fig.3b-3c and Fig.3e-3f). These results notably highlight that the $T_c$ value of the given material is influenced not only by its superfluid density ($\rho_s$) but also by the conductivity ($\sigma$) at $T_c$. In other words, the $T_c$ value of a given superconductor is closely linked to the ratio of $\rho_s/\sigma$, which serves as a combined response factor that reflects the intrinsic electromagnetic state of the superconducting system to external electromagnetic fields[12,35]. To provide a clearer representation of the relationship between the relevant physical quantities governing $T_c$, we reformulate Homes' law as $T_c = (1/A) \times (\rho_s/\sigma)$, where $T_c$ is proportional to the ratio of the response coefficients of magnetic field ($\rho_s$) to electric field ($\sigma$). This modified formulation may provide a plausible explanation for why $T_c$ remains insensitive to external pressure while presents dramatic changes in $\rho_s(P)$ and $\sigma(P)$.

Finally, we propose that the dome-like evolution of $\rho_s$ and $\sigma$ with pressure may be attributed to the interplay between the effects of hole-doping and pressure on the metallization process and the subsequent development of superconductivity. It is expected that our results obtained from the compressed high-$T_c$ superconductors and the reformulated Homes' law ($T_c = (1/A) \times (\rho_s/\sigma)$) may provide valuable insights for a better understanding on the mechanism of the high-$T_c$ superconductors, a longstanding challenge of condensed matter physics and material sciences.

These authors with star (*) contributed equally to this work.
Correspondence and requests for materials should be addressed to Liling Sun (llsun@iphy.ac.cn or liling.sun@hpstar.ac.cn)

**Acknowledgements**

The work in China was supported by the National Key Research and Development Program of China (Grant No. 2021YFA1401800 and 2022YFA1403900), the NSF of China (Grant Numbers Grants No. U2032214, 12122414, 12104487 and 12004419) and the Strategic Priority Research Program (B) of the Chinese Academy of Sciences

(Grant No. XDB25000000). The work at BNL was supported by the US Department of Energy, office of Basic Energy Sciences, contract no. DOE-sc0012704.

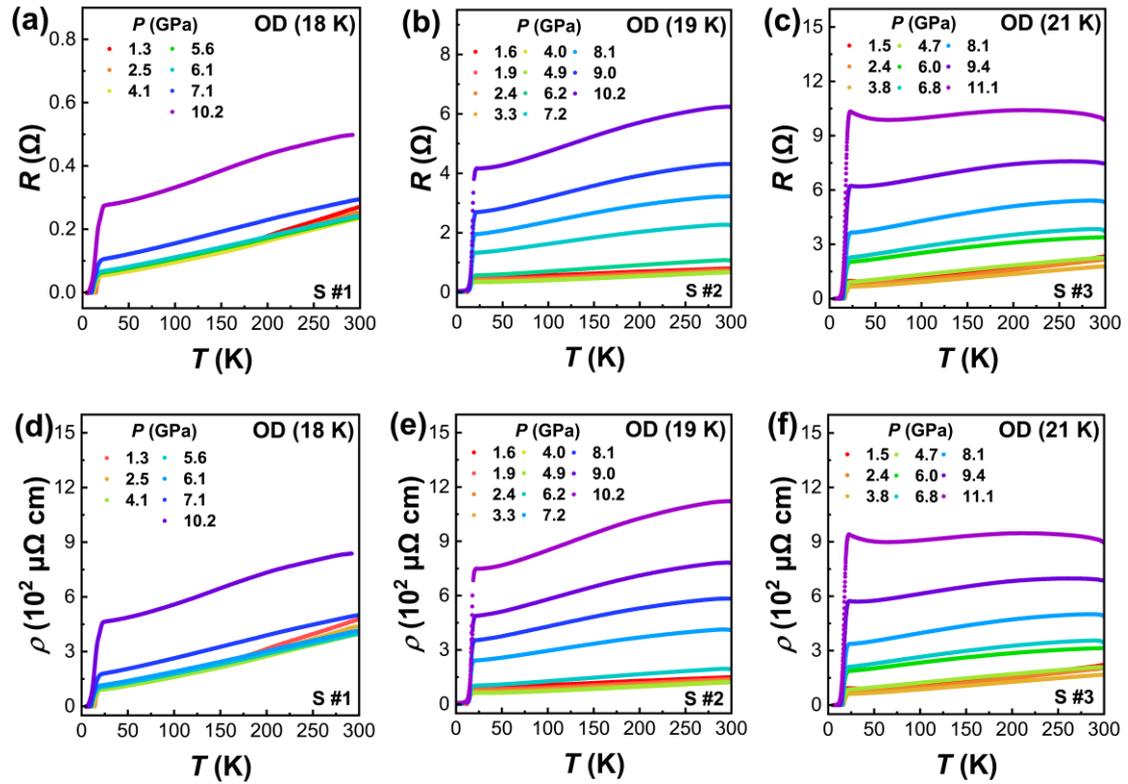

**Figure 1 The transport properties of the over doped $Bi_{1.74}Pb_{0.38}Sr_{1.88}CuO_{6+\delta}$ (Bi-2201) superconductors.** (a)-(c) Temperature dependence of the electrical resistance at different pressures for the three samples with superconducting transition temperature ($T_c$) of 18 K, 19 K and 21K, respectively. (d-f) Resistivity versus temperature for the same three samples. OD stands for the over-doped samples.

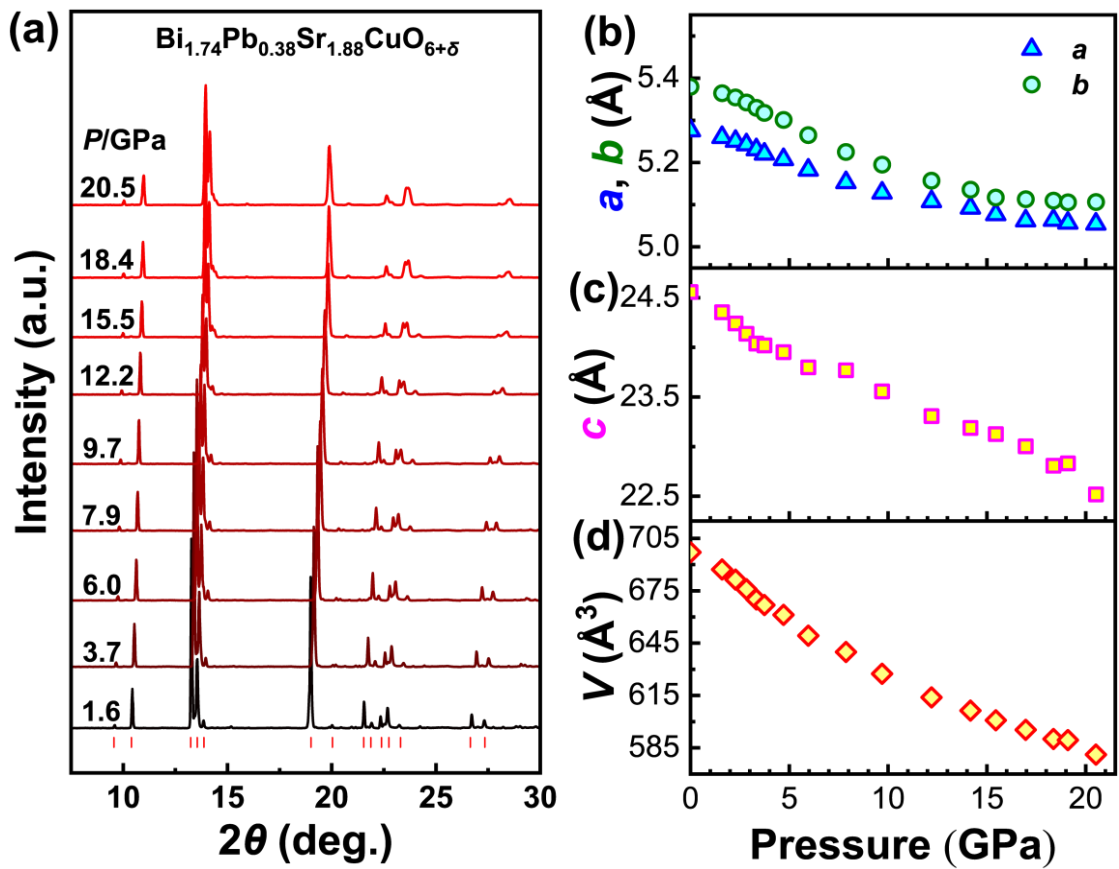

**Figure 2 High-pressure structural information of the $Bi_{1.74}Pb_{0.38}Sr_{1.88}CuO_{6+\delta}$ (Bi-2201) superconductor at room temperature.** (a) X-ray diffraction patterns collected at different pressures. All peaks can be indexed well by the orthorhombic structure in space group *Pnan*. (b) and (c) Pressure dependences of lattice parameters (*a*, *b* and *c*). (d) Volume (*V*) as a function of pressure. All the data indicates that no structural phase transition occurs in the pressure range investigated.

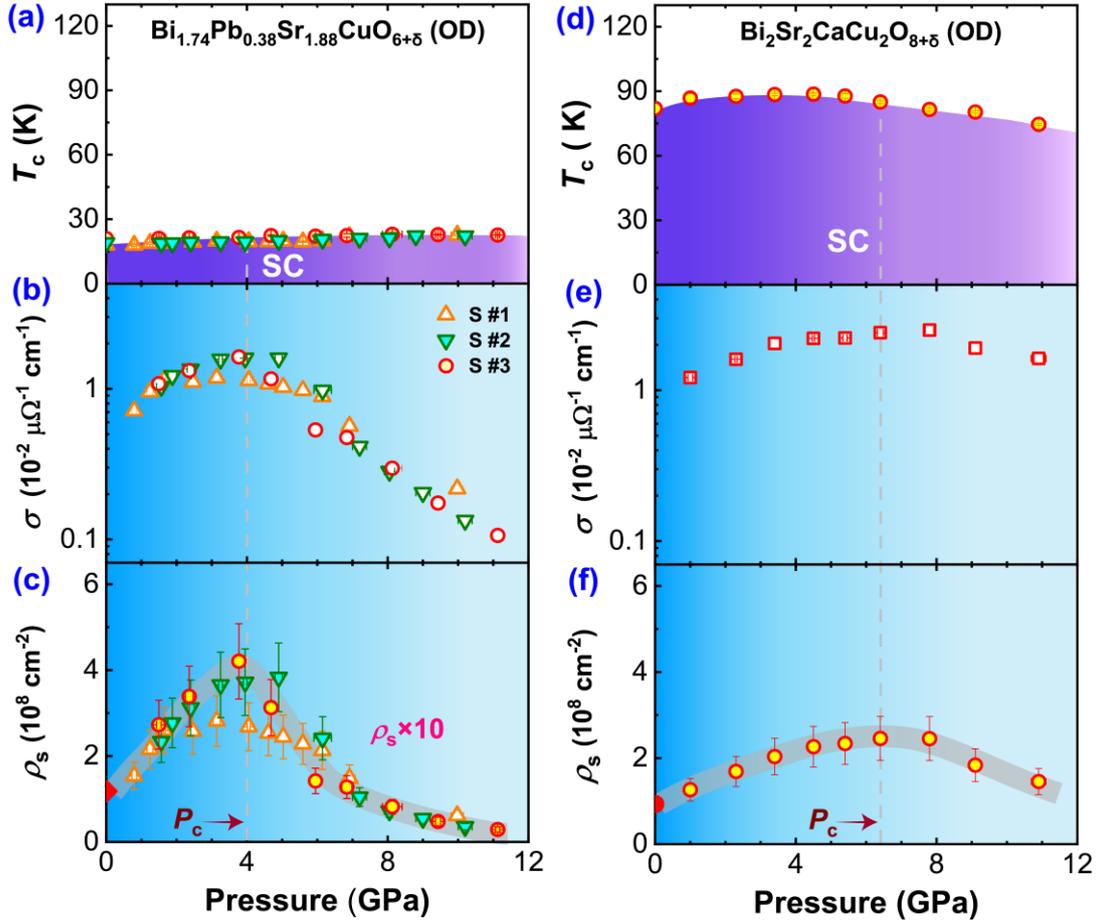

**Figure 3 Summary of the superconducting transition temperature ($T_c$), conductivity ($\sigma$) and superfluid density ($\rho_s$) as a function of pressure for the over doped $Bi_{1.74}Pb_{0.38}Sr_{1.88}CuO_{6+\delta}$ (Bi-2201) and slightly over doped $Bi_2Sr_2CaCu_2O_{8+\delta}$ (Bi-2212) superconductors.** (a) and (d) Pressure-$T_c$ phase diagram for the Bi-2201 and Bi-2212 superconductors. The data displayed in Fig. 3(d) are taken from Ref. 36. (b) and (e) Pressure dependence of $\sigma$ for the Bi-2201 and Bi-2212 superconductors. The data displayed in Fig. 3(e) are extracted from the results reported in Ref. 36. (c) and (f) Plots of $\rho_s$ (red circle, orange and green triangles) versus pressure. The red diamond and solid represent the ambient-pressure superfluid density taken from the Ref. 30 and Ref. 31 respectively, the values of which are consistent with the ones extrapolated from our high-pressure data.